\documentstyle[epsfig,epsf,eqsecnum,aps,amssymb,multicol]{revtex}

\begin{document}
\draft \preprint{HEP/123-qed}
\title{Peak effect and dynamic melting of vortex matter in NbSe$_2$ crystals}
\author{N. Kokubo, K. Kadowaki, and K. Takita\\
}
\address{Institute of Materials Science, University of Tsukuba, 1-1-1, Tennoudai, Tsukuba,
Ibaraki 305-8573, Japan}

\date{\today}
\maketitle
\begin{abstract}
We present a mode locking (ML) phenomenon of vortex matter
observed around the peak effect regime of 2H-NbSe$_2$ pure single
crystals. The ML features allow us not only to trace how the shear
rigidity of driven vortices persists on approaching the second
critical field, but also to demonstrate a dynamic melting
transition of driven vortices at a given velocity. We observe the
velocity dependent melting signatures in the peak effect regime,
which reveal a crossover between the disorder-induced transition
at small velocity and the thermally induced transition at large
velocity. This uncovers the relationship between the peak effect
and the thermal melting.
\end{abstract}

\begin{multicols}{2}
\narrowtext
\renewcommand{\theequation}{\arabic{equation}}

%\section*{1. Introduction}
Discontinuous jumps in equilibrium magnetization observed well
below the mean field line in clean high $T_c$ cuprate
superconductors \cite{MeltingZeldov}, have been widely recognized
as a hallmark of the thermodynamic melting transition (MT), which
separates a vortex solid state, where elastic interaction
dominates and quasi-long ranged, crystalline correlations develop
in vortex structure, from a vortex liquid state in which thermal
fluctuations disrupt the crystalline order and the shear rigidity
vanishes.

The melting signature is often accompanied by the pronounced peak
anomaly of the magnetization or critical current \cite{Shi}, known
as the peak effect (PE), originating from the rapid softening of
the vortex lattice and the random pinning potential due to
disorder quenched in a host material. The close proximity of the
MT on the PE has led to a reexamination of the physical properties
close to the second critical field $H_{c2}$, especially on low
$T_c$ materials like Nb \cite{NeutGammel,NeutLing,NeutForgan} and
NbSe$_2$ \cite{PeakBhatt,Troyanovski}. Various experimental
results evidence the pinning-induced structural transformation
into a disordered array of vortices, indicating the dominant
influence of the quenched disorder in the PE regime
\cite{NeutGammel,Troyanovski,Corbino,Marchevsky}. However, the
presence of the MT and its relation with the PE remain
controversial \cite{NeutLing,NeutForgan}.

The effect of the quenched disorder changes remarkably when
vortices are driven by a transport current.  As proposed
theoretically \cite{Kosheleve,MovingGrass,Scheidl} and
experimentally \cite{PeakBhatt,Yaron,MoGeIV}, at low drive the
"shaking action" of the random pinning due to the quenched
disorder disrupts largely the internal periodicity in moving
structure of driven vortices, while at high drive the influence of
the disorder diminishes and the elastic interaction becomes
dominant, resulting in moving solid states at large velocity.

This leads to an unique opportunity to study whether the "driven
lattice" undergoes thermally induced MT and also whether the
transition occurs in the disorder dominant, PE regime.
%This would allow us to separate the peak effect
%from the melting transition.
%If the disorder induces little effect at high velocity, it is
%interesting to ask whether the "driven lattice" melts thermally at
%the thermodynamic melting point.
%%"driven lattice" melts thermally at the thermodynamic melting
%point.
In this letter, we present experimental evidences for the
\emph{dynamic melting transition}(DMT) of driven lattice observed
just above the PE of NbSe$_2$ by means of mode locking (ML)
experiment. ML is a dynamic synchronization between rf drive
superimposed on dc drive and collective lattice (elastic) modes
excited over driven vortices at internal frequencies
$f_{int}=qv/a$ with the average velocity $v$, the lattice spacing
$a$ and an integer $q$ \cite{kkbprlprb,RutdynamicMelting}. When
the internal frequency and the external drive frequency are
harmonically related, i.e. $f_{int}=pf$ or $v=(p/q)fa$ with an
integer $p$, the elastic modes are dynamically locked on rf drive.
We observe this dynamic resonance as multiple current steps in a
current-voltage ($IV$) curve or multiple peaks in a differential
conductance ($dI/dV$) curve.  The ML features allow us not only to
trace how the shear rigidity of driven vortices persists on
approaching $H_{c2}$, but also to demonstrate the DMT of driven
vortices at a given velocity. We will show that the velocity
dependence of dynamic melting signatures unveils the relationship
between the PE and the MT.

%\begin{figure}
%\begin{center}
%\epsfig{file=Figure1.eps, width=8cm} \vspace{0.1cm} \label{Fig.1}
%\caption{(a) A peak anomaly of critical current $I_c$ measured by
%varying magnetic field $H$ at 4.2K. Peak field $H_p$ is denoted by
%an arrow. A dotted line marks the dynamic melting field measured
%at high velocity. The inset displays how differential resistance
%$R_d$ measured at a large dc current of 30 mA $(\gg I_c)$ depends
%on field. See the text for definition of the second critical field
%$H_{c2}$.  (b) The phase diagram of NbSe$_2$. Both $H_{c2}$
%and $H_p$ are plotted. See the text for arrows.}
%\end{center}
%\end{figure}

%\section*{2. Experimental}
Platelets of 2H-NbSe$_2$ pure single crystal used in this study
were grown by an iodine vapor transport method \cite{Takita}. The
thinnest three crystals (typically 10$\mu$m thick) were cut in the
bar shape and cleaved with no significant optical surface damage.
Contacts for the four-probe method were made by indium solder. The
crystals have the superconducting transition temperature $T_c$ of
7.2 K determined from the mid-point of resistive transition. The
transition width is about 50 mK between 10$\%$ and 90$\%$ of the
normal state resistance just above $T_c$ (e.g., R(7.4K)). Typical
values of residual resistance ratio $R$(295K)/$R$(7.4K) for the
crystals are ranged between 30 and 40. However, no significant
change of the room temperature resistivity $\rho_0$ of $1.2 \times
10^{-6} \Omega$m was observed in crystals from different batches
\cite{Takita}. This allows us to estimate the thickness of the
cleaved crystals from the room temperature resistance $R$(295K).

%\begin{figure}
%\begin{center}
%\epsfig{file=Figure2.eps, width=7cm} \vspace{0.1cm} \label{Fig.2}
%\caption{A mode locking (ML) phenomenon of driven vortices. (a)
%Differential conductance ($dI/dV$)-voltage ($V$) curves measured
%in 1.90 T with superimposing 3 MHz rf current of various
%amplitudes. For clarity, the origin for each curve is shifted
%vertically. ML conditions are denoted. (b) Frequency $f$
%dependence of ML voltages for the fundamental and higher
%harmonics. (c) Rf current $I_{rf}$ vs. the fundamental ML current
%width $\Delta I$ obtained by integrating the fundamental ML peak
%shown in (a) with respect to the flux flow base line (see the
%inset). See text for a broken curve. (d) $\Delta I$ for 3 MHz vs
%$I_{rf}$ observed in 1.995 T just below the dynamic melting
%point.}
%\end{center}
%\end{figure}

The measurements discussed in this study were carried out on a
cleaved crystal with a dimension of 0.72mm($l$)$\times$
0.63mm($w$)$\times$ 0.9($t)\mu$m.  Figure 1(b) shows results of
$H_{c2}$ on the crystal obtained from a measurement of flux flow
resistance with magnetic field applied perpendicularly to the
platelet. As displayed in the inset to Fig. 1(a), $H_{c2}$ is
determined from the intersectional point between the linear
extrapolation of flow resistance indicated by a solid line and the
normal resistance by a dotted line \cite{Berghuis}. The slope of
$H_{c2}$ on $T$ at $T_c$ $(\equiv -S)$ is -0.72 T/K, in good
agreement with magnetization results
\cite{Troyanovski,Takita,Kobayashi}. Shown in Fig. 1(a) is a plot
of the critical current $I_c$ vs magnetic field $H$, obtained by
ramping field up (and also down) after cooling the crystal to 4.2
K in the zero field (indicated by arrows in Fig. 1(b))
\cite{Helium}. Here, $I_c$ is determined from $IV$ curves by a
10$\mu V/$m criterion. It exhibits a peak behavior below
$\mu_0H_{c2}$ of 2.14 T, characterized by the peak field $H_p$ of
1.93 T.  We note that, because of inhomogeneous edges in the strip
geometry, $I_c$ \emph{below} $H_p$ has both bulk and edge
contributions \cite{SurfaceBarrier,EdgeBulk}, resulting in
broadening for the lower part of the PE.

%Moreover, the argument by the driving force $F(=IB)$ would lead to
%incorrect picture for vortex dynamics. Thus, instead of $F$, in
%the study, we use the vortex average velocity, determined from the
%ML velocity condition of $v=(p/q)fa$.

%\section*{3. Result and Discussion}

Figure 2(a) shows a series of $dI/dV-V$ curves measured in 1.90 T
(just below $H_p$) with superimposing 3 MHz rf current of various
amplitudes \cite{Penetrationdepth}. Application of a relatively
large rf current of 16.7 mA, for instance, induces clear
conductance ML peaks at equidistant voltages. First peak
corresponds to the fundamental ($p/q=1/1$) and others are higher
harmonics ($2/1$, and $3/1$). Small peaks at sub-harmonic
conditions with $q>1$ are also detected. As expected in the
velocity condition ($v=(p/q)fa$), those resonance voltages
$(\propto v)$ increase linearly with rf drive frequency (see Fig.
2(b)). Thus, by changing $f$, we can detect the shear rigidity of
driven vortices at various velocities.

In Fig. 2(c), we show how the fundamental resonance depends on rf
current $I_{rf}$. Here, the fundamental ML current width $\Delta
I$ is obtained by integrating the conductance ML peak with respect
to the flux-flow-base line (see the inset to Fig. 2(c)). It shows
an oscillatory behavior with $I_{rf}$, which agrees qualitatively
with squared Bessel function of the first kind displayed by a
broken curve. This behavior is expected when the random pinning
due to the quenched disorder excites elastic modes in driven
elastic lattices \cite{kkbprlprb}.  In other frequencies and
fields, similar oscillatory behavior of $\Delta I$ on $I_{rf}$ are
observed in different rf current scales. For simplicity, we focus
mainly on the first maximum $\Delta I_{max}$ of the fundamental ML
width in the following analysis.

Next we turn to the magnetic field dependence of the ML resonance.
In Fig. 3(a), we show how the fundamental ML voltage $V_{1/1}$
evolves with $H$. As observed, $V_{1/1}$ for 7 MHz increases with
$H$ with downward curvature. This is in good quantitative
agreement with the fundamental ML-voltage condition displayed by a
solid curve; $V_{1/1}=Af\Phi_0l/a_0$ with the equilibrium lattice
spacing $a_0(=\sqrt{A\Phi_0/B})$, a factor $A(=2/\sqrt{3})$, the
sample length $l$(=0.72 mm), and the vortex density $B=\mu_0H$.
Note that no fitting parameter is used for this comparison.

As magnetic field approaches $\mu_0H_{c2}$(=2.14 T), however, the
ML resonance suddenly disappears at a certain field.  An example
is displayed in Fig. 3(b), where the $dI/dV$ curves measured with
superimposing a 7 MHz rf current in different fields are shown.
The fundamental ML peak is clearly observed in lower fields. As
$H$ is ramped up, the peak becomes small and it seems to vanish
around $\mu_0H$=2.01 T, above which the $dI/dV$ curves become
featureless.

%\begin{figure}
%\begin{center}
%\epsfig{file=Figure3.eps, width=6cm} \vspace{0.1cm} \label{Fig.3}
%\caption{(a) Magnetic field dependence of the fundamental ML
%voltage $V_{1/1}$ measured with superimposing a 7 MHz rf current.
%A solid curve represents the fundamental ML-voltage condition (see
%text). (b) The differential conductance curves measured with a 7
%MHz rf current in fields from 1.96 T to 2.02 T by a constant field
%step of 0.01 T. For clarity the origin for each curve is
%vertically shifted.}
%\end{center}
%\end{figure}

The characteristics are better displayed by plotting $\Delta
I_{max}$ against field, which is given in Fig. 4(a). $\Delta
I_{max}$ for 3 MHz, for instance, starts to show a rapid drop
around 1.98T (defined as the onset field $H_{on}$) and disappears
around 2.00T, while it is insensitive to $H$ below $H_{on}$. We
define the vanishing field $H^*$ from the linear extrapolation to
$\Delta I_{max}=0$ as indicated by a solid line. Above this field,
no ML resonance appears at any amplitude of $I_{rf}$, indicating
the absence of the shear rigidity in driven vortices. Thus, the
vanishing field should mark the DMT from coherent to liquid like
incoherent flow states \cite{kkbprlprb,RutdynamicMelting}. We note
that this phenomenon emerges just above the PE where $I_c$ drops,
and therefore it should be driven by the thermal fluctuations, in
addition to the quenched disorder.

The reduction of the disorder effect is expected in the ML
features measured at high frequencies (large velocities). As shown
in Fig. 4(a), $\Delta I_{max}$ measured at a high frequency of 9
MHz exhibits a sharper drop at slightly higher fields. Namely,
both the onset and the vanishing fields increase slightly, while
the field range where the linear drop of the ML width occurs
becomes narrow. We observe even sharper phenomenon at high
frequencies, indicating the sudden disappearance of the shear
rigidity in driven vortices at large velocities.
%These features are reminiscent of results in
%YBCO crystals in which melting signatures become sharp on
%approaching the middle of the melting line from the critical
%points \cite{NishizakiReview}.

%Those ML phenomena appear within $\sim$10 $\%$ below the second
%critical field of $H_{c2}(=2.1T)$, where the shear rigidity of the
%driven vortex lattice becomes small due to the overlap between
%neighboring vortices cores.  Thus, in addition to the thermal
%fluctuations, the quenched disorder may induce considerable
%plasticity in driven motion, which may be responsible for the ML
%features.  As the velocity of driven vortices is increased by
%driving force, the plasticity induced by the quenched disorder is
%effectively reduced. Thus, systematic studies on how those ML
%features vary with velocity would provide insight into not only
%the effect on quenched disorder on vortex dynamics, but also the
%disorder free, thermal melting phenomenon at the high velocity
%limit.

The striking frequency (velocity) dependence is also observed in
the onset and vanishing fields, characterizing the dynamic melting
signatures. In Fig. 4(b) results of $H^*$ and $H_{on}$ are plotted
against velocity determined from the fundamental velocity
condition of $v=fa_0$.
%$=f\sqrt{(\sqrt{3}/2)\Phi_0/B}$ with $B=\mu_0H$.
As observed, $H^*$ (open symbols) increases with $v$ and exhibits
a saturation behavior toward $\approx$2.01 T at larger velocities.
$H_{on}$ (solid symbols) also exhibits a similar saturation
behavior toward the nearly identical field at larger velocities.
The difference between $H^*$ and $H_{on}$ at a constant velocity
decreases with increasing $v$ and becomes negligible
experimentally at large velocity. These behaviors imply that at
small velocity the influence of the disorder facilitates the
occurrence of the DMT, while at large velocity it becomes
negligible and the thermal effect dominates on the phenomenon.
Thus, the velocity dependence of the dynamic melting signatures
reveals a crossover between the disorder- and the thermally-driven
transitions.

%Moreover, the difference between the vanishing and the onset
%fields becomes narrow with increasing velocity.  These results
%indicate that,

%\begin{figure}
%\begin{center}
%\epsfig{file=Figure4.eps, width=7cm} \vspace{0.1cm} \label{Fig.4}
%\caption{(a) Magnetic field vs. the maximized fundamental ML
%current width $\Delta I_{max}$ for 3 MHz and 9 MHz. For clarity,
%the origin for 9 MHz data is shifted vertically. Onset $H_{on}$
%and vanishing $H^*$ fields are indicated by arrows. Their velocity
%dependencies are given in (b). A broken line represents the
%saturation field. See text for solid and dotted curves.}
%\end{center}
%\end{figure}

Regarding the vanishing field at a given velocity as a
crystallization velocity $v_c$ at the field, the saturation
behavior is qualitatively similar to the dynamic ordering picture
proposed by Koshelev and Vinokur \cite{Kosheleve}, in which
combined influences of the thermal fluctuations and the quenched
disorder on vortex dynamics near the thermodynamic melting
transition are taken into account. In this model, the
crystallization velocity diverges as $v_c=v_0/(1-T/T_M)$ on
approaching the melting point $T_M$ from below.
%Although only the
%temperature driven situation is proposed, this is also applicable
%to the field driven one. Indeed, in recent ML results on amorphous
%MoGe films \cite{ML_MOGE}\cite{channel}, temperature and field
%driven $v_c$ data exhibit similar diverging behavior toward the
%melting points of $T_M$ and $H_M$, respectively.
As displayed by a solid curve, the $H^*$ data are well
approximated by a similar function of $v_c=v_0/(1-H/H_M)^\alpha$
with the melting field $\mu_0H_M=2.01$T (indicated by a broken
line), $\alpha=$1 and $v_0=0.33$mm/s over nearly two decades of
velocity \cite{OnsetH}.
%We also see similar agreement of the $H_{on}$ data ($>1.9$ T) with
%the identical function by the same melting field $\mu_0H_M=2.01$T
%and a large value of by $v_0=1.4$mm/s.
This allow us to identify the saturation field as the
thermodynamic melting point for the vortex lattice. Thus, the
sharp drop of ML by the saturation field signifies the thermally
induced, DMT of driven lattice.
%We note
%that the influence of the applied current on the thermodynamic
%quantities (e.g. $H_{c2}$) is negligible since the maximum total
%current density ($j_{rf}+j_{dc}$) applied to the crystal is less
%than $10^8$ A/m$^2$, which is four decades smaller than the
%theoretical depairing current density.

Let us compare the saturation field to a recent quantitative
theoretical result for the thermodynamic melting line proposed by
Li and Rosenstein \cite{LiRosen}. The MT is set in when the lowest
Landau-level-scaled temperature $a_T=-(\sqrt{N_{Gi}/2}\pi t
h)^{-2/3}(1-t-h)=-9.5$, with their Ginsburg number
$N_{Gi}=((k_BT_c\gamma/4\pi^2 \mu_0H_c(0)^2\xi^3)^2)/2$, a reduced
temperature $t = T/T_c$, and a reduced field $h=H/H_{c2}(0)$.
Using the anisotropy parameter $\gamma=3$ \cite{Kobayashi}, the
Ginzburg-Landau (GL) thermodynamic field at $T=$0
$\mu_0H_c(0)=0.24$ T\cite{Kobayashi}, and the GL second critical
field $\mu_0H_{c2}(0)=ST_c=5.18$ T for the present crystal, we
find that the saturation field of 2.01 T at 4.2 K gives $a_T=$-10.
A nearly similar result of $a_T$=-9 is also found in other crystal
with $\mu_0H_M$=2.02 T at 4.2 K. These are in good agreement with
the theoretical result.

Next, we discuss flow states near the DMT. In the regime between
the onset and vanishing fields, an anomalous dependence of $\Delta
I$ on $I_{rf}$ is observed, of which example is shown in Fig.
2(d). There is a minimum rf current to observe ML, indicating the
influence of the quenched disorder disrupting the shear rigidity
in driven vortices.
%Thus, in addition to dc current, rf current plays an important
%role for the flow ordering. This is strikingly different from a
%liquid like incoherent flow state above $H^*$ where no ML feature
%occurs at any large amplitudes of $I_{rf}$.
Such threshold behavior in ML favors with a smectic flow (or a
layered liquid) state \cite{Balents,MLKolton}, in which only the
periodicity perpendicular to the flow direction is preserved.
Thus, the $H^*$ curve would mark a dynamic transition between the
smectic and the liquid flow states. Meanwhile, for $H<H_{on}$, the
threshold rf current for ML disappears and $\Delta I$ shows the
the Bessel function like oscillatory behavior on $I_{rf}$,
suggesting a moving solid state. Thus, the solid-smectic
transition would occur at the $H_{on}$ curve. Similar two step
process of the DMT has been proposed theoretically
\cite{MovingGrass,Scheidl}.

Finally, we compare the melting curve of $H^*$ to the peak
behavior of $I_c$.
%Next, we turn to the behavior of the depinning force in the
%relevant field region. In Fig. 1(a), a plot of the pinning force
%density $F_p$ vs. a reduced field $b (=H/H_{c2})$ is given. Here
%$F_P=BJ_c=BI_c/wt$ in which $I_c$ is determined from a dc $I-V$
%curve by an electric field criterion of 2$\mu$V/cm. As observed,
%the peak effect of the depinning force appears clearly near the
%saturation field ($b_M$=0.958) indicated by a broken line. The
%onset and peak reduced fields of the peak effect are 0.84 and
%0.92, respectively.
%As indicated by a dotted line in Fig. 1(a), the dynamic melting
%point (or $H_M$) at high velocity coincides with neither the onset
%nor the peak fields of the peak behavior. It emerges in the higher
%part of the PE close to $H_{c2}$ . Meanwhile, on reducing
%velocity, $H^*$ approaches $H_p$ and seems to disappear (or be
%below our experimental resolutions) around 1.9 T just below $H_p$.
%A recent study on edge and bulk transport \cite{EdgeBulk} reveals
%that the bulk component of $I_c$ begins to show a peak anomaly
%just below the peak temperature of $I_c$.
As indicated by a dotted line in Fig. 1(a), the dynamic melting
point (or $H_M$) at large velocity emerges in the higher part of
the PE. On reducing velocity, $H^*$ approaches $H_p$ and seems to
disappear (or be below our experimental resolutions) around 1.9 T
just below $H_p$. A recent study on edge and bulk transport
\cite{EdgeBulk} reveals that the bulk component of $I_c$ begins to
show a peak anomaly just below the peak temperature of $I_c$. This
prompts us to relate the disappearance of $H^*$ to the disordering
transition for the PE. Thus, the $H^*$ curve would be a locus of
the melting point connecting smoothly the PE with the MT as a
function of the disorder effect (velocity).
%We note that the appearance of the
%thermodynamic melting point in the higher part of the peak effect
%is consistent with the recent result of a fast transport study
%\cite{XiaoSpinodal}.

%Scanning tunnelling microscopy experiment on NbSe$_2$ crystals
%has observed a transition from the collective vortex motion to
%positional fluctuations of individual vortices around the onset
%of the peak effect \cite{Troyanovski}. we believe that, in the
%equilibrium, the full development of plasticity done by the
%quenched disorder results in an amorphous arrangement of vortices
%prior to the melting point.

%In summary, we present the ML phenomenon of the vortex matter near
%the peak effect of 2H-NbSe$_2$ single crystals. At large velocity,
%a very sharp drop of the ML feature occurs at the velocity
%independent field, signifying the thermally induced dynamic
%melting transition of driven lattice. In addition, the transition
%occurs in the higher part of the peak effect in the critical
%current. From those findings, we conclude that, when driven
%enough, the pinning induced disordering transition is suppressed,
%instead, signatures of thermally induced melting phenomenon
%recovers in driven state.

In summary, employing ML techniques, we have presented
experimental evidences for the DMT of driven vortex lattice
observed just above the PE of 2H-NbSe$_2$ single crystals. At
small velocity the transition driven predominantly by disorder
occurs near $H_p$, while at large velocity the transition by
thermal fluctuations is observed by $H_M (>H_p)$. The velocity
dependence of the melting signatures reveals the smooth crossover
between the PE and the MT. We hope that this work stimulate

%We hope that the present ML techniques unveil various
%intrinsic properties of periodic media hindered by extrinsic
%origins.

%\section*{Acknowledgements}
N. K. used facilities in the cryogenic center in the university of
Tsukuba. N. K. thanks P. H. Kes, A. E. Koshelev and T. Nishizaki
for useful comments. This work was partly supported by the grant
in Aid for Scientific research (Grant No. 16710063) from the
Ministry of Education, Science and Culture, Japan.
% Bibliographic references with the natbib package:
% Parenthetical: \citep{Bai92} produces (Bailyn 1992).
% Textual: \citet{Bai95} produces Bailyn et al. (1995).
% An affix and part of a reference:
%   \citep[e.g.][Ch. 2]{Bar76}
%   produces (e.g. Barnes et al. 1976, Ch. 2).

\end{multicols}

\newpage
\section*{Figure caption}
Fig. 1 (a) Peak anomaly of critical current $I_c$ measured by
varying magnetic field $H$ at 4.2K. The peak field $H_p$ is
denoted by an arrow. A dotted line marks the dynamic melting field
measured at high velocity. The inset displays how differential
resistance $R_d$ measured at a large dc current of 30 mA $(\gg
I_c)$ depends on field. The definition of the second critical
field $H_{c2}$ is given in text. (b) Phase diagram of NbSe$_2$.
Both $H_{c2}$ and $H_p$ are plotted. See text for arrows.

Fig. 2  Mode locking (ML) phenomenon of driven vortices. (a)
Differential conductance ($dI/dV$)-voltage ($V$) curves measured
in 1.90 T with superimposing 3 MHz rf current of various
amplitudes. For clarity, the origin for each curve is shifted
vertically. ML conditions are denoted. (b) Frequency $f$
dependence of ML voltages for the fundamental and higher
harmonics. (c) Rf current $I_{rf}$ vs. the fundamental ML current
width $\Delta I$ obtained by integrating the fundamental ML peak
shown in (a) with respect to the flux-flow base line (see the
inset). See text for a broken curve. (d) $\Delta I$ for 3 MHz vs
$I_{rf}$ observed in 1.995 T between $H^*$ and $H_{on}$.

Fig. 3 (a) Magnetic field dependence of the fundamental ML voltage
$V_{1/1}$ measured with superimposing a 7 MHz rf current. A solid
curve represents the fundamental ML-voltage condition (see text).
(b) Differential conductance curves measured with a 7 MHz rf
current in fields from 1.96 T to 2.02 T by a constant field step
of 0.01 T. For clarity the origin for each curve is vertically
shifted.

Fig. 4 (a) Magnetic field vs. the maximized fundamental ML current
width $\Delta I_{max}$ for 3 MHz and 9 MHz. For clarity, the
origin for 9 MHz data is shifted vertically. Onset $H_{on}$ and
vanishing $H^*$ fields are indicated by arrows. Their velocity
dependencies are given in (b). A broken line represents the
saturation field. See text for solid and dotted curves.

%\begin{figure}
%\begin{center}
%\epsfig{file=figure1.eps, width=12cm} \vspace{0.1cm} \label{Fig.1}
%\caption{}
%\end{center}
%\end{figure}

%\begin{figure}
%\begin{center}
%\epsfig{file=figure2.eps, width=12cm} \vspace{0.1cm} \label{Fig.2}
%\caption{}
%\end{center}
%\end{figure}

%\begin{figure}
%\begin{center}
%\epsfig{file=figure3.eps, width=12cm} \vspace{0.1cm} \label{Fig.3}
%\caption{}
%\end{center}
%\end{figure}

%\begin{figure}
%\begin{center}
%\epsfig{file=figure4.eps, width=12cm} \vspace{0.1cm} \label{Fig.4}
%\caption{}
%\end{center}
%\end{figure}


\begin{thebibliography}{00}
\bibitem{Helium} During ML experiments, the crystals were immersed
in liquid $^4$He to aviod heating.
\bibitem{Penetrationdepth} Flux-flow-skin depth defined
as $\sqrt{2\rho_f/2\pi f\mu_0}$ with flux flow resistivity
$\rho_f$ is 15 $\mu$m at 3MHz and 1.9T.
\bibitem{OnsetH}We also see similar agreement of the $H_{on}$ data ($>1.9$ T) with
the identical function by $\mu_0H_M=2.01$T and a large value of by
$v_0=1.4$mm/s diplayed by a broken curve.
\end{thebibliography}
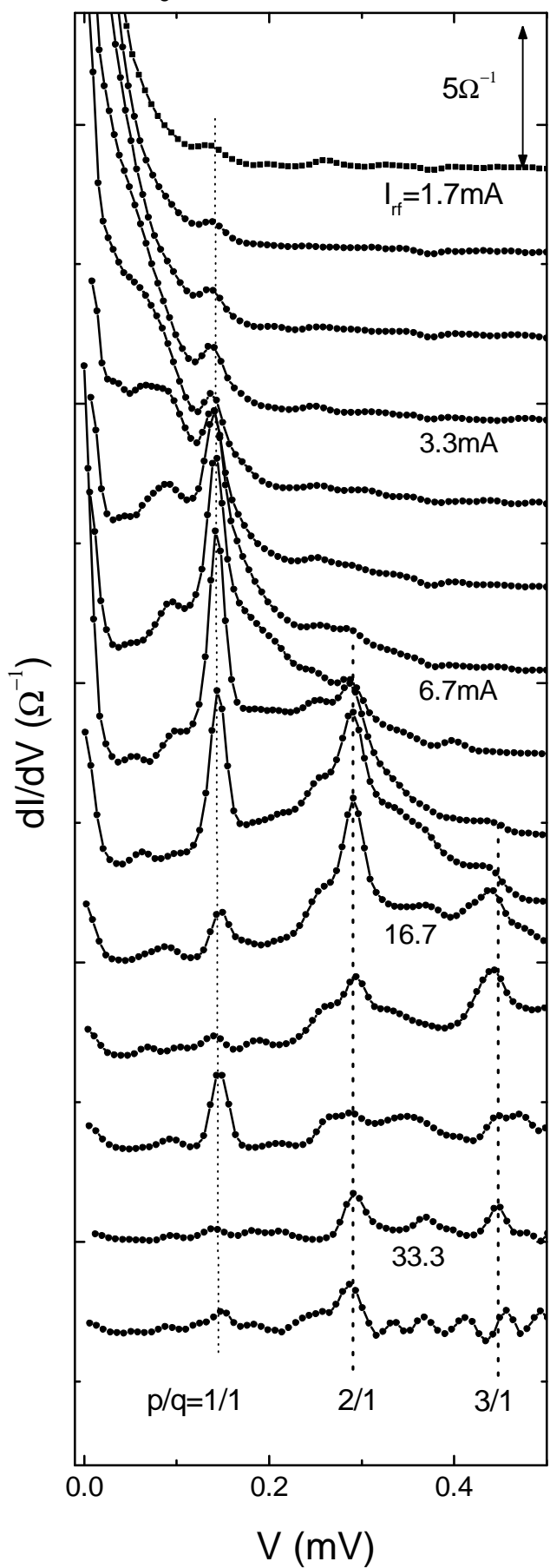
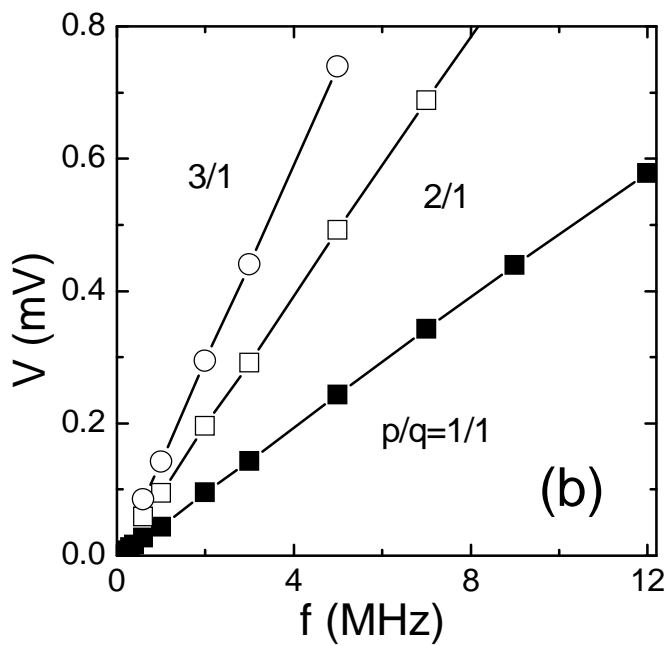
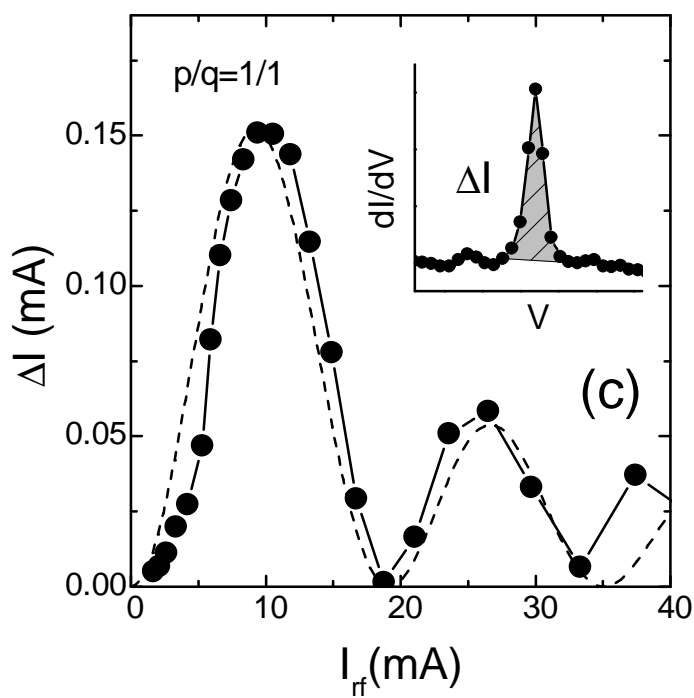
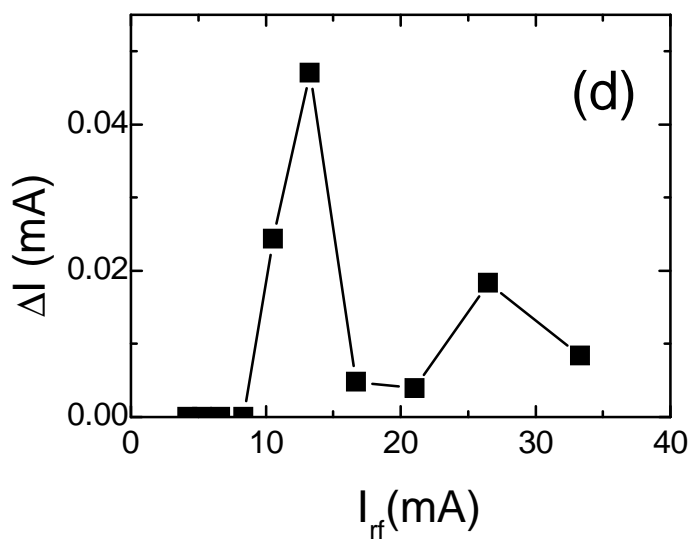
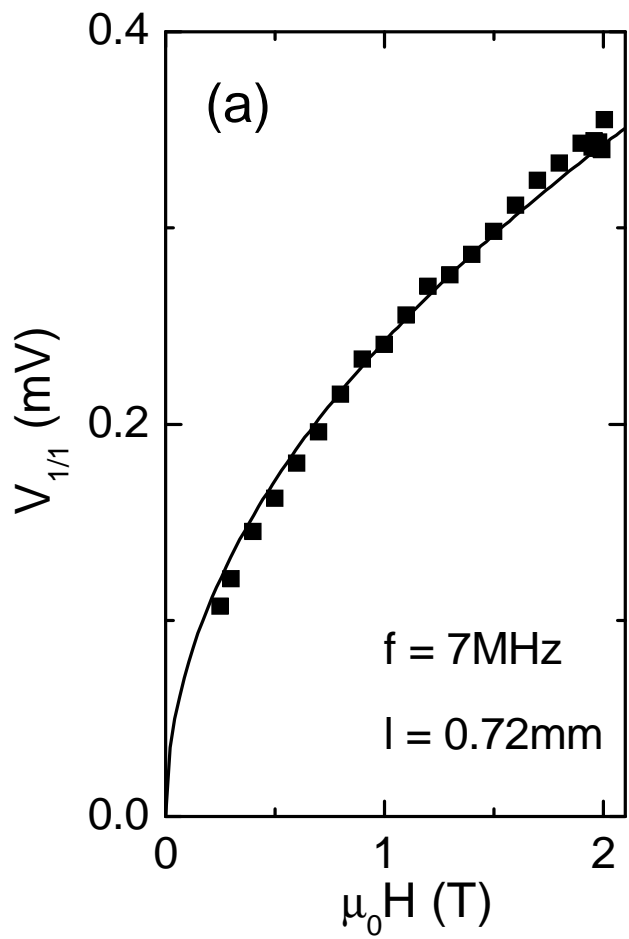
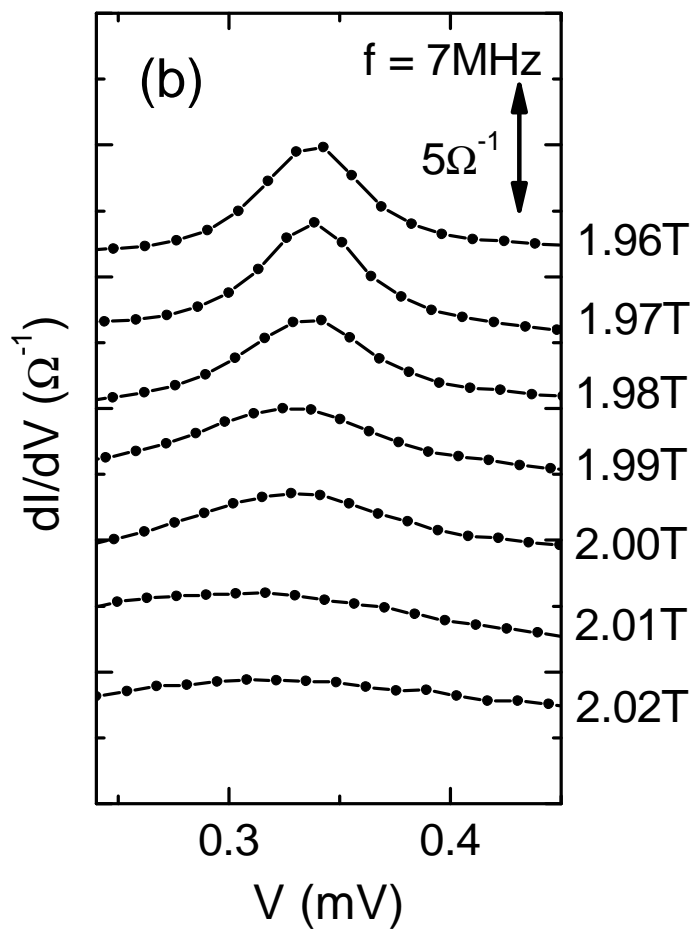
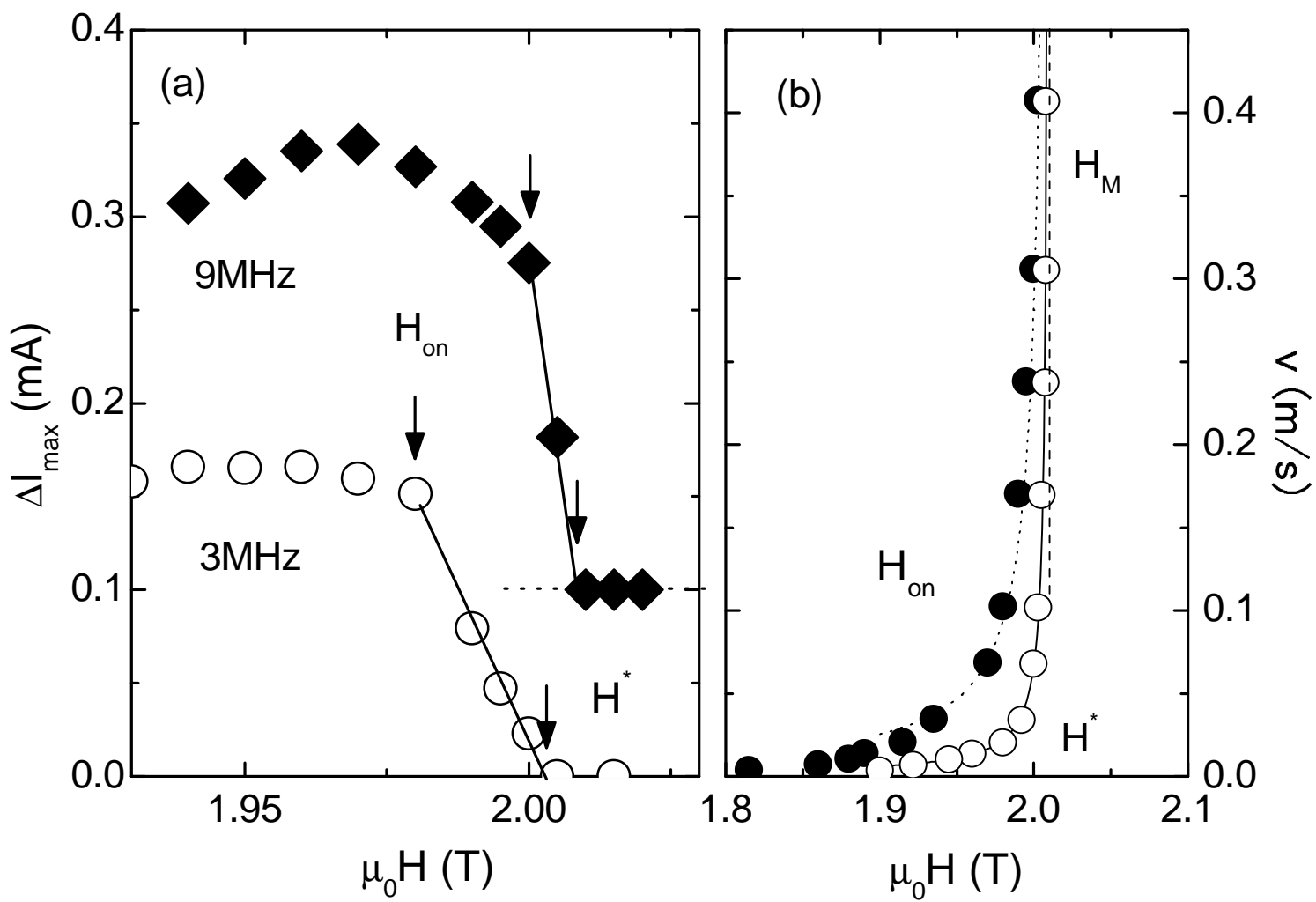
\end{document}